\documentclass[  10pt, showkeys, superscriptaddress]{revtex4}
\usepackage[utf8]{inputenc}
\usepackage{graphics,epsfig}

\usepackage{graphicx}
\usepackage{hyperref} 
\usepackage{dcolumn}
\usepackage{amsmath}
\newcommand{\be}{\begin{equation}}
\newcommand{\ee}{\end{equation}}
\newcommand{\bea}{\begin{eqnarray}}
\newcommand{\eea}{\end{eqnarray}}
\newcommand{\nn}{\nonumber}
\begin{document}

\title{\Large  $4D$ AdS Einstein-Gauss-Bonnet black hole with Yang-Mills field and its thermodynamics }
\author{Dharm Veer Singh} 
\email{veerdsingh@gmail.com}
\affiliation{Department of Physics, Institute of Applied Science and Humanities, GLA University, Mathura, 281406 India.}
\author{Benoy Kumar  Singh}
\email{benoy.singh@gla.ac.in}
\affiliation{Department of Physics, Institute of Applied Science and Humanities, GLA University, Mathura, 281406 India.}
\author{Sudhaker Upadhyay}
\email{sudhakerupadhyay@gmail.com}
\affiliation{Department of Physics, K. L. S. College,  Nawada, Bihar 805110, India}
\affiliation{Department of Physics, Magadh University, Bodh Gaya,
 Bihar  824234, India}
\affiliation{Inter-University Centre for Astronomy and Astrophysics (IUCAA), Pune, \\Maharashtra 411007, India}
\affiliation{School of Physics, Damghan University, P.O. Box 3671641167,\\ Damghan, Iran}

\begin{abstract}
\noindent We derive an exact black hole solution for the  Einstein-Gauss-Bonnet gravity  with Yang-Mills field in  $4D$ AdS spacetime and investigate its thermodynamic properties to calculate exact expressions for the black hole mass, temperature, entropy and heat capacity. The  thermodynamic quantities get modification in the presence of Yang-Mills field, however, entropy remains unaffected by the Yang-Mills charge. The solution exhibits $P\--v$   criticality  and belongs to the universality class of Van der Waals fluid. We study the effect of Gauss-Bonnet coupling and Yang-Mills charge on the critical behaviour and black hole phase transition. We observe that the values of  critical exponents  increase with the Yang-Mills charge and decrease with the Gauss-Bonnet coupling constant.
\end{abstract} 
\maketitle

\section{Introduction}
 Einstein theory of general relativity (GR)   is one of the most 
 successful theories in theoretical physics. Despite its appreciable 
 achievements,
there are still some unsolved problems in the universe such as   the hierarchy problem, the
cosmological constant problem, and the late time accelerated expansion of the Universe. This reflects  that GR is not
the ultimate theory and needs further generalization.
One of the possible generalization for the standard Einstein-Hilbert action is  by adding the higher curvature
terms. Natural candidates for the higher curvature corrections are provided by Lovelock
gravities, which are the unique theories that give rise to generally covariant field equations  \cite{lav}. 
Gauss-Bonnet gravity, also called as Einstein-Gauss-Bonnet (EGB), is the simplest extension of the Einstein-Hilbert action to include
higher curvature Lovelock terms. A Gauss-Bonnet term   appears naturally  in the low energy effective action of
string theory \cite{zwe}.

Thermodynamic properties of black holes have been studied for many years \cite{hen, sud4,sud5,sud6,sud7} and found
that black hole spacetime can not only be assigned standard thermodynamic variables such as temperature or
entropy, but were also shown to possess rich phase structures and admit critical phenomena. Hawking and Page \cite{haw} were  the first who studied the thermodynamic properties of AdS black holes and found the existence of
a certain phase transition in the phase space of the (nonrotating uncharged) Schwarzschild-AdS black hole. This was the first order phase transition between thermal  AdS space and the Schwarzschild-AdS black hole, with the latter becoming
thermodynamically preferred   above a certain critical temperature.  The study of the phase transitions in
higher curvature gravity are subject of present research \cite{05,06,07}.
For instance,   Van der Waals behaviour, (multiple)-rentrant phase transitions, critical points  and
isolated critical points for a variety of asymptotically AdS black holes  can be found in  Refs. \cite{08,09,10,11,12,13}.
Although there   various black hole  solutions exist, but the black hole 
solution  for the EGB gravity with Yang-Mills field and their thermal 
properties remain  unstudied. This provides us an opportunity to bridge this gap. 

Recently, Glavan and Lin \cite{gla} introduced the 4-dimensional theory of gravity with Gauss-Bonnet correction by   re-scaling the Gauss-Bonnet coupling $\alpha\to\alpha/D-4$. 
 Here,   it is important to note
 that taking the limit $D \rightarrow 4$ after $\alpha\to\alpha/D-4$ either breaks part of diffeomorphism or leads to extra gravitational degrees
of freedom. Clearly, both of these violate conditions of the Lovelock's theorem \cite{lav}. A consistent theory
of $D \rightarrow 4$ EGB gravity with two dynamical degrees of freedom is proposed  by breaking the
temporal diffeomorphism invariance  \cite{Ao1, Ao2}.
 The generalization to other spherically symmetric  black holes has also discussed in Refs. \cite{fran,Singh:2020nwo,Hennigar:2020lsl}. Other probes  include studies
of the  regular black holes \cite{Singh:2020mty,Yang:2020jno,Kumar:2020uyz,32,33,35,36,37}, quasi-normal modes \cite{Devi:2020uac,Mishra:2020gce}, gravitational lensing \cite{Jin:2020emq,Islam:2020xmy}, thermodynamic properties, $P-v$ criticality and phase transition \cite{Konoplya:2020cbv,Wei:2020poh,Singh:2020xju,Zhang:2020obn, Singh20, Singh21, hendi19,hendi20,hendi18, hendi17, hendi16,he,hendi2017}.

In this work, we first obtain a static spherically symmetric black hole 
solution for EGB gravity with Yang-Mills field. The solution by Glavan and Lin can be
recovered   from this generalized solution. Actually, the resulting solution interpolates with the Glavan and Lin solution  \cite{gla} in the Yang-Mills charge. In order to  analyse thermodynamic property, we estimate the mass, Hawking temperature from area-law and entropy which satisfy the first-law of thermodynamics.  The stability of the system is studied by computing the heat capacity.  Here, we observe that  the  heat capacity  curve  is discontinuous at the critical radius. Further, we notice that there is a flip of sign in the heat capacity around critical radius. Thus,   we  can  say that  $4D$  EGB AdS  Yang-Mills black holes whose horizon radius is smaller than the critical radius are    thermodynamically stable, otherwise it is thermodynamically   unstable.  Thus, there is a phase transition occurs at critical horizon radius.   The  phase transition occurs from the higher to lower mass black holes. The critical radius  increases along with   parameter Gauss-Bonnet coupling constant. The   large value of Yang-Mills charge   makes the black hole stable. Moreover, we compute the critical exponents,  a universal property of phase transition, which exactly  match with the mean field theory. Finally, we study the  Gibbs free energy  which analyses the phase transition of the black holes   analogues with the Van der Waals phase transition and find that   Yang-Mills charge  $\nu$ and  Gauss-Bonnet coupling constant $\alpha$ affect the Gibbs free energy of the system.

The paper is organized in seven sections. In section \ref{sec2}, we obtain a black hole solution for the   EGB  action in the presence of  Yang-Mills field. Section \ref{sec3} outlines the first law of thermodynamics and the stability of black hole. The black hole is considered as  the  Van der Waals  fluid in section \ref{sec4}. Here, critical behaviour of the model and behaviour of Gibbs free energy are 
calculated in this section. The values of critical exponents are estimated in section \ref{sec5}.  The discussions and final comments are made in the last section. 

\section{Black Hole Solution for Einstein-Gauss-Bonnet gravity in  $ 4D$ AdS specetime}\label{sec2}The  Einstein-Gauss-Bonnet (EGB) gravity recently formulated in four dimensions ($4D$) by Glavan and Lin \cite{gla} as of higher dimension field equation by rescaling the coupling constant. The consistent realization of the idea suggested in Refs. \cite{Ao1, Ao2} either breaks the part of  diffeomorphism or leads to extra  gravitational degrees of freedom. A  consistent realization of $4D$ EGB gravity which is invariant under spatial diffeomorphism and breaks the time diffeomorphism  is proposed by the Aoki, Gorji and Mukhhyama (AGM) \cite{Ao1, Ao2}.
In order to get consistent theory of $4D$ Einstein-Gauss-Bonnet gravity, we should start with Arnowitt-Deser-Misner (ADM) metric \cite{Ao1, Ao2}
\be
ds^2=-n^2dt^2+\gamma_{ij}(dx^i+n^jdt)(dx^j+n^jdt),
\label{admmetric}
\ee
where $n$, $n_i$ and 
$\gamma_{ij}$ correspond to lapse function, the shift
vector and the spatial metric, respectively.

The action for EGB gravity  theory (in natural unit) is given by \cite{Ao2}
\be
S=\frac{1}{16\pi }\int dt d^3x n\sqrt{\gamma}\left[2{ R}-{\cal M}+ {\alpha}  R^2_{GB}\right].\label{2}
\ee
Here $R$ is the Ricci scalar of the   ``spatial metric", $\alpha$ is the Gauss-Bonnet coupling coefficient $\alpha\geq 0$ and  Gauss-Bonnet Lagrangian  is defined as
\be
R^2_{GB}=\frac{1}{2}\left[8{ R}^2-4{  R}{\cal M}-{\cal M}^2\right]-\frac{8}{3}\left[8{ R}_{ij}{ R}^{ij}-4{ R}_{ij}{\cal M}^{ij}-{\cal M}_{ij}{\cal M}^{ij}\right],
\ee
where ${\cal M}$ is the trace of ${\cal M}_{ij}$ defined as
\be
{\cal M}_{ij}={  R}_{ij}+{\cal K}_{\ k}^k {\cal K}_{ij}-{\cal K}_{ik}{\cal K}^k_{\ j} \quad \text{with}\quad {\cal K}_{ij}=\frac{1}{2n}\left[\dot{\gamma}_{ij}-2D_{(i}{n}_{j)}-\gamma_{ij}D^2\lambda_{GF}\right].
\ee
 Here, $R$, $R_{ij}$ and ${\cal K}_{ij}$  are the Ricci scalar, Ricci tensor and extrinsic curvature respectively,  the covariant derivative compatible with the
spatial metric is denoted by $D_i$ and Lagrange multiplier   related to the gauge-fixing constraint is denoted by $\lambda_{GF}$.  One of the properties of the consistent AGM theory described by the
action (\ref{2}) is that the $d\rightarrow 3$ limit of the $d + 1$-dimensional solution of EGB gravity is the
solution of the AGM theory if it has vanishing Weyl tensor of the spatial metric in $d + 1$
dimensions.

Since, our main objective here is to study a black hole solution with Yang-Mills field and their 
thermal properties in  $4D$ EGB gravity.  Therefore, it is worth writing the Yang-Mills Lagrangian as
\be
{\cal F}_{YM}=- F_{\mu\nu}^{(a)}F^{(a)\mu\nu},
\ee
where field-strength tensor is defined in terms of gauge potential ${\cal A}_{\mu}^{(a)}$ as $F_{\mu\nu}^{(a)}=2\nabla_{[\mu}{\cal A}_{\nu]}^{(a)}+f_{(b)(c)}^{(a)}{\cal A}_{\mu}^{(b)}{\cal A}_{\nu}^{(c)}$, with  the structure constant   $f_{(b)(c)}^{(a)}$ which  can be calculated with  the  generator of  gauge group $SU(2)$.

In order to obtain  static spherically symmetric black hole solution of the EGB gravity with Yang-Mills field (given in Appendix \ref{ap}), we write the following line element:
\begin{equation}
ds^2 = -f(r)dt^2+\frac{1}{f(r)} dr^2 + r^2 d\Omega_{D-2},
\label{metric}
\end{equation}
where $d\Omega_{D-2}$ is the metric of a $(D-2)$-dimensional   sphere $S^{D-2}$.  Comparing this metric with the ADM metric (\ref{admmetric}), one confirms that
\begin{eqnarray}
 n^2=f(r),\qquad  n^i=0,  \qquad  \text{and}  \qquad 
 \gamma_{ij}=\mbox{diag}\left(\frac{1}{f(r)}, r^2, r^2\sin^2\theta,\ldots\right).
\end{eqnarray}

\begin{figure*}[ht]
\begin{tabular}{c c c c}
\includegraphics[width=.5\linewidth]{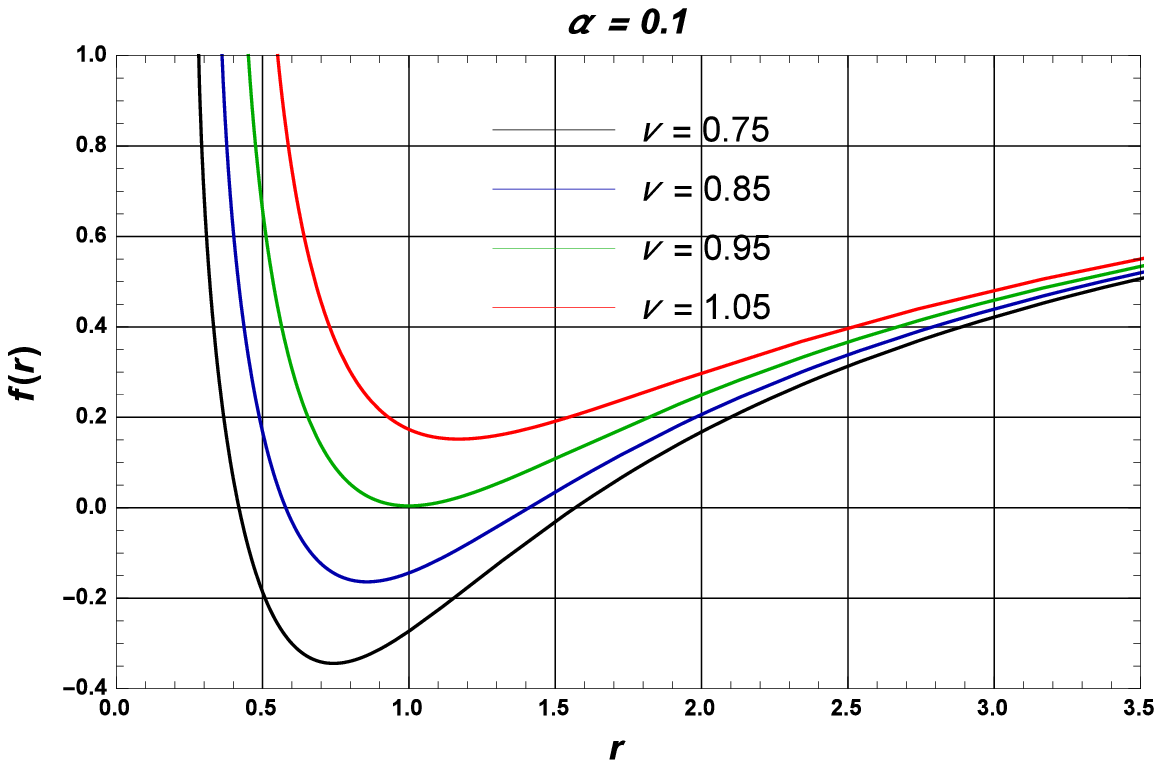}
\includegraphics[width=.5\linewidth]{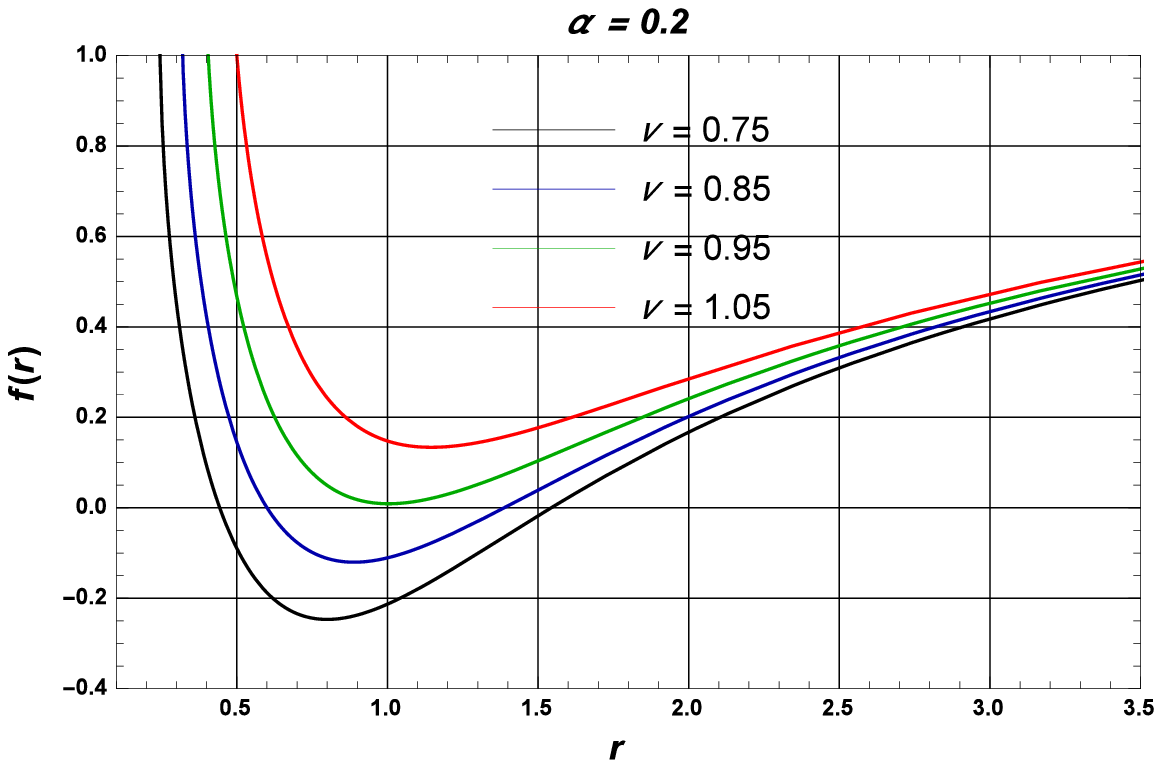}
\end{tabular}
\caption{The plots of $f(r)$ vs $r$ for different values of Yang-Mills charge $\nu$ for  $-ve$ branch of solution (\ref{sol1})  with fixed value of mass $(M=1)$ and ${l=20}$ for different Gauss-Bonnet coupling parameter: $\alpha=0.1$ (left) and  $\alpha=0.2$ (right). }
\label{fig:1}
\end{figure*}

In order to specify  $f(r)$,   we appoint the position dependent generators ${\cal T}_{(r)}$, ${\cal T}_{(\theta)}$ and ${\cal T}_{(\phi)}$ of the gauge group rather the standard one $t_{(1)}$, $t_{(2)}$ and $t_{(3)}$. The relations between the basis of $SU(2)$ group and the standard basis are given as
\begin{eqnarray}
&&{\cal T}_{(r)}=\sin\theta\cos\nu\phi t_{(1)}+\sin\theta\sin\nu\phi t_{(2)}+\cos\theta t_{(3)},\\
&&{\cal T}_{(\theta)}=\cos\theta\cos\nu\phi t_{(1)}+\cos\theta\sin\nu\phi t_{(2)}-\sin\theta t_{(3)},\\
&&{\cal T}_{(\phi)}=-\sin\nu\phi t_{(1)}+\cos\nu\phi t_{(2)}.
\end{eqnarray}
These generators satisfy the following commutation relations:
\begin{eqnarray}
\left[{\cal T}_{(r)},{\cal T}_{(\theta)}\right]={\cal T}_{(\phi)},\qquad\qquad \left[{\cal T}_{(\phi)},{\cal T}_{(r)}\right]={\cal T}_{(\theta)},\qquad\qquad \left[{\cal T}_{(\theta)},{\cal T}_{(\phi)}\right]={\cal T}_{(r)}.
\end{eqnarray}
Meanwhile, we consider the Wu-Yang choice of the gauge potential having the following 
non-zero components \cite{bal}:
\begin{eqnarray}
{\cal A}_{\theta}^{(a)}=\delta_{(\phi)}^{(a)} \quad \text{and}\qquad {\cal A}_{\phi}^{(a)}=-\nu\sin\theta\delta_{(\theta)}^{(a)}.
\end{eqnarray}
The Wu-Yang ansatz and the Yang-Mills tensor field lead to the following non-vanishing component of Yang-Mills field  $F_{\theta\phi}^{(r)}=\nu \sin\theta$. The $(r,r)$ components of the Eq. (\ref{3}) in the limit $D\to 4$ becomes
\begin{equation}
r^5-2r^3\alpha(f(r)-1)f'(r)+r^4(f(r)-1)+r^2\alpha (f(r)-1)^2+\Lambda r^2+\frac{\nu}{r^2}=0.
\label{rr}
\end{equation}
 For $D \to 4$, we have following solution 
 \begin{eqnarray}
f_{\pm}(r)=1+\frac{r^2}{2\alpha}\left(1\pm\sqrt{1+4\alpha\left(\frac{2M}{r^3}-\frac{\nu^2}{r^4}-\frac{1}{l^2}\right)}\,\right),
\label{sol1}
\end{eqnarray}
where it is natural to adopt $\Lambda$ in terms scale length $l$ as  {\bf $\Lambda=-3/l^2$} for AdS solution.
This is an exact solution corresponds to the two branch of solution depending on the choice of signature. Here, it is worth mentioning that although one can define the local notions like trapped surfaces and
so on, defining the global quantities like event horizon is ambiguous.
 The $+ ve$ signature is not physical as it reduces to Reissner--Nordstr\"om black hole with negative mass and imaginary charge. However, $- ve$ signature reduces to the Schwarzschild solution for $\alpha\to 0$, $\nu\to 0$ and $l\to \infty$. Therefore, we limit  ourselves to the $- ve$ branch of solution throughout the manuscript.  In this solution, $M$ is the integration constant related to the mass of the black hole, $\alpha$ is the Gauss-Bonnet coupling  and $\nu$ is Yang-Mills charge related to the hair.  In the limit of $\nu=0$,  the solution 
(\ref{sol1}) reduces to the $4D$ EGB AdS black hole solution
\begin{eqnarray}
f (r)=1+\frac{r^2}{2\alpha}\left(1-\sqrt{1+4\alpha\left(\frac{2M}{r^3}-\frac{1}{l^2}\right)}\,\right).\label{sol11}
\end{eqnarray}
{The solution by Glavan and Lin  \cite{gla}  can be
recovered from this solution in the   limit of  $l\to \infty$. Also, the solution (\ref{sol1})  reduces  to   AdS  Schwarzschild  black hole when $\nu=0$ and $\alpha\to 0$.} This describes   AdS  Schwarzschild Yang-Mills black hole when $\alpha \to 0$ with following metric:
\be
f(r)=1-\frac{2M}{r}+\frac{\nu^2}{r^2}+\frac{r^2}{l^2}.
\ee
As we know that  $f(r)=0$ gives the horizons of the black hole, but  Eq. (\ref{sol1}) is a transcendental equation  and therefore can not be solved analytically. So, we plot  the graph  to estimate the horizons as shown in the Fig. \ref{fig:1}. 
\begin{center}
	\begin{table}[h]
		\begin{center}
			\begin{tabular}{l l r l| r l r l r}
				\hline
				\hline
				\multicolumn{1}{c}{ }&\multicolumn{1}{c}{ $\alpha=0.1$  }&\multicolumn{1}{c}{}&\multicolumn{1}{c|}{ \,\,\,\,\,\, }&\multicolumn{1}{c}{ }&\multicolumn{1}{c}{}&\multicolumn{1}{c}{ $\alpha=0.2$ }&\multicolumn{1}{c}{}\,\,\,\,\,\,\\
				\hline
				\multicolumn{1}{c}{ \it{$\nu$}} & \multicolumn{1}{c}{ $r_-$ } & \multicolumn{1}{c}{ $r_+$ }& \multicolumn{1}{c|}{$\delta$}&\multicolumn{1}{c}{$\nu$}& \multicolumn{1}{c}{$r_-$} &\multicolumn{1}{c}{$r_+$} &\multicolumn{1}{c}{$\delta$}   \\
				\hline

				\,\,\, 0.75\,\,& \,\,0.419\,\, &\,\,  1.567\,\,& \,\,1.146\,\,&0.70&
				\,\, 0.443\,\,&\,\,1.543\,\,&\,\,1.100\,\,
				\\
				\
				\,\, 0.85\,\, & \,\,0.579\,\, &\,\, 1.409\,\,& \,\,0.830\,\,&0.80&
				\,\, 0.600\,\,&\,\,1.388\,\,&\,\,0.788\,\,
				\\
				\,\,\, 0.95$(\nu_{c})$\,\, &  \,\,0.995\,\,  &\,\,0.995\,\,&\,\,0
				\,\,&0.90$(\nu_{c})$& \,\, 0.995\,\,&\,\,0.995\,\,&\,\,0\,\,
				\\

				\hline
				\hline
			\end{tabular}
		\end{center}
		\caption{Inner  horizon ($r_-$), outer horizon ($r_+$) and $\delta=r_+-r_-$ for different values of Yang-Mills charge $\nu$ with fixed value of mass $(M=1)$ and AdS length scale ${l=20}$.}
				\label{f3}
	\end{table}
\end{center}
In table \ref{f3}, we identify the different values of the parameters.
Here, $r_{+}$ corresponds to the outer horizon while $r_{-}$ represents the inner horizon.  Elementary analysis of the zeros of $f(r)=0$ reveals that it has no zeros if  $\nu > \nu_c$, one double zero
 if $\nu = \nu_c$, and two simple zeros if
 $\nu < \nu_c$, (Fig. \ref{fig:1}). From the figure, it is evident that the two horizons coincide at the critical radius
  \begin{equation}\label{Rc}
 r_c= \sqrt{\frac{l^2\sqrt{1+\frac{12}{l^2}(\alpha+\nu^2)}-l^2}{6}}.
\end{equation}
These cases, therefore, describe   $4D$ EGB Yang-Mills black hole  with degenerate horizon and a  non-extreme
black hole with both outer and inner Killing horizons,  respectively.  It is clear that the critical value of $M_c$ and $r_c$ depend upon the parameters $\alpha$ and $\nu$.   Also, the radius of the outer horizon increases  with decrease in the Yang-Mills charge  $\nu$  and the  Gauss-Bonnet coupling $\alpha$ as shown in Fig. \ref{fig:1} and Table \ref{f3}. 

From the standard relation $f(r_+)=0$, the mass parameter of  $4D$ EGB AdS Yang-Mills black hole is calculated as
\begin{equation}
M_+=\frac{r_+}{2}\left(\frac{r_+^2+\alpha+\nu^2}{r_+^2}+\frac{r_+^2}{l^2}\right).
\end{equation}
From this expression, we can study various thermodynamical quantities of the system. 
\section{Thermodynamics and stability of   black hole}\label{sec3}
In this section, we compute essential thermodynamic quantities which help in checking first-law of thermodynamics and the stability of black holes. 
In particular we derive Hawking temperature, entropy and  heat capacity, chronologically. 

The Hawking temperature associated with the black hole is defined in term of horizon radius as following:
\begin{equation}
T_+=\frac{\kappa}{2\pi}=\frac{1}{4\pi}f'(r)|_{r=r_+},
\end{equation}
where parameter $\kappa$ refers to the surface gravity. 

 In our case, the above definition leads to the following value for temperature:
\begin{equation}
T_+=\frac{1}{4\pi r_+}\left(\frac{r_+^2-\alpha-\nu^2}{r_+^2+2\alpha}+\frac{3r_+^4}{l^2 (r_+^2+2\alpha)}\right).
\label{temp1}
\end{equation}
For vanishing Yang-Mills charge $\nu=0$, this temperature (\ref{temp1}) reduces to the case of the $4D$ AdS EGB black hole,
\begin{equation}
T_+=\frac{1}{4\pi r_+}\left(\frac{r_+^2-\alpha}{r_+^2+2\alpha}+\frac{3r_+^4}{l^2(r_+^2+2\alpha)}\right).
\end{equation}
The relation (\ref{temp1}) also reduces to the case of    AdS   Schwarzschild Yang-Mills black hole  and    AdS   Schwarzschild black hole for $\alpha =0$ and $\nu=0, \alpha = 0$, respectively.

\begin{figure*}[ht]
\begin{tabular}{c c c c}
\includegraphics[width=.5\linewidth]{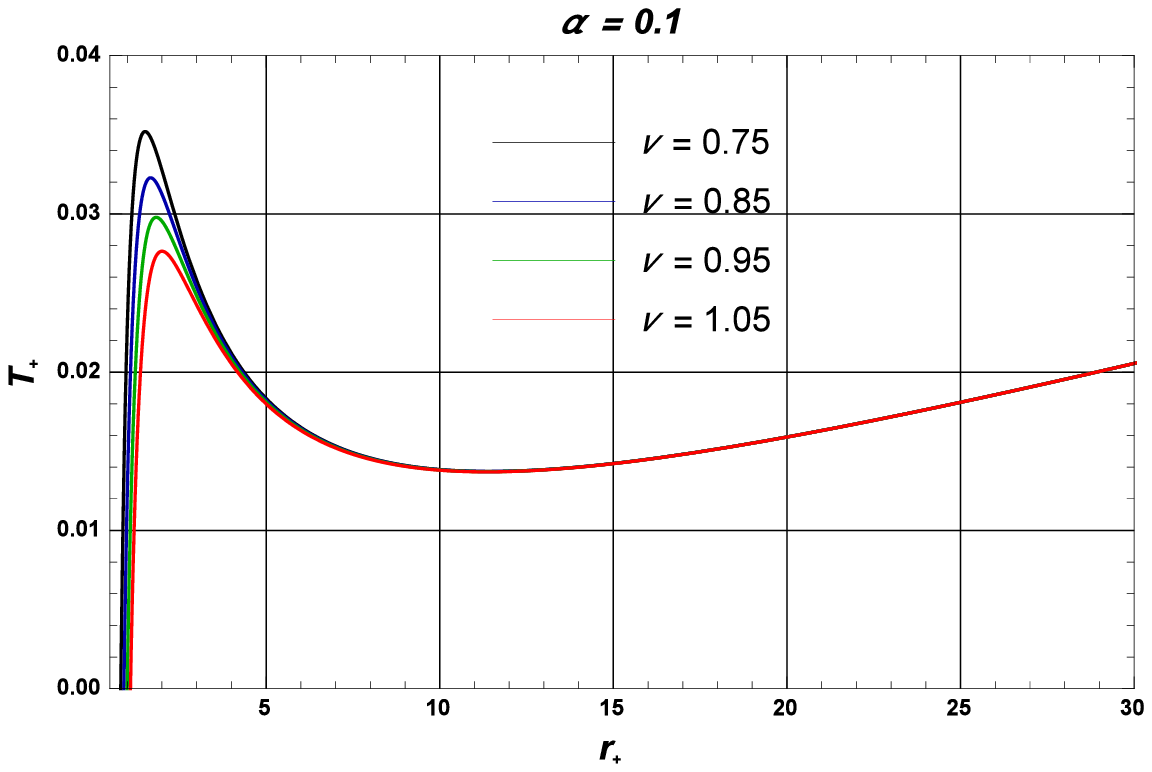}
\includegraphics[width=.5\linewidth]{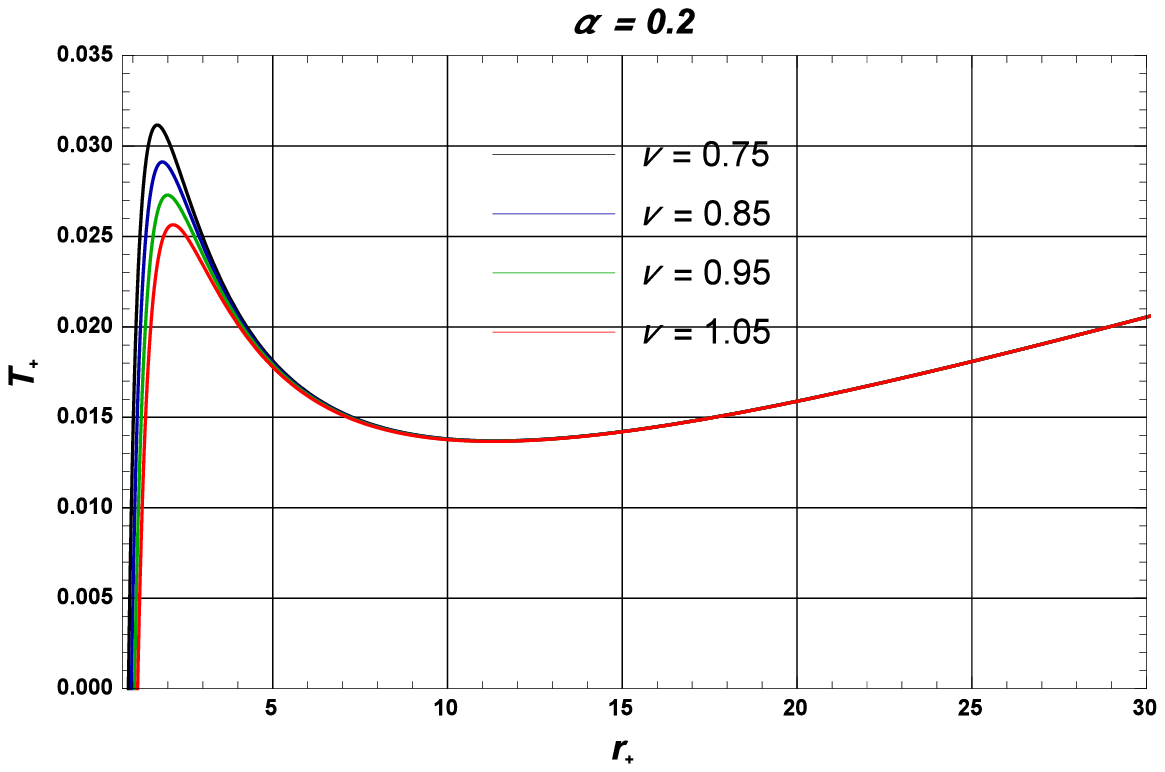}
\end{tabular}
\caption{The plot of temperature $T$ vs $r_+$ for different values of Yang-Mills charge   $\nu$ for Gauss-Bonnet coupling constant $\alpha=0.1$ (left) and $\alpha=0.2$ (right) with fixed value of $l=20$. }
\label{fig:2}
\end{figure*}

Now, in  order to study the behaviour of temperature, we plot temperature with respect to event horizon for different values of Yang-Mills charge  as shown in Fig. \ref{fig:2}.
From the figure, this is obvious that the Hawking  temperature of the $4D$ EGB AdS Yang-Mills black hole  grows to a maximum (say $T_{max}$) and  then drops to minimum temperature.  It turns out that the maximum value of the Hawking temperature decreases with increase of the Yang-Mills charge  $\nu$ and Gauss-Bonnet coupling constant  $\alpha$.  The  Hawking temperature vanishes  at the critical radius shown in Table \ref{tab:temp}.
\begin{center}
	\begin{table}[h]
		\begin{center}
			\begin{tabular}{l|l r l r l| r l r r r}
				\hline
				\hline
				\multicolumn{1}{c}{ }&\multicolumn{1}{c}{ }&\multicolumn{1}{c}{ }&\multicolumn{1}{c}{$\alpha=0.1$  }&\multicolumn{1}{c}{ \,\,\,\,\,\, }&\multicolumn{1}{c|}{ }&\multicolumn{1}{c}{  }&\multicolumn{1}{c}{ }&\multicolumn{1}{c}{$\alpha=0.2$}\,\,\,\,\,\,\\
				\hline
				\multicolumn{1}{c|}{$\nu$} &\multicolumn{1}{c}{ } &\multicolumn{1}{c}{ 0.75 } & \multicolumn{1}{c}{ 0.85 }& \multicolumn{1}{c}{0.95}& \multicolumn{1}{c|}{1.05} &\multicolumn{1}{c}{}&\multicolumn{1}{c}{0.75} &\multicolumn{1}{c}{ 0.85}   & \multicolumn{1}{c}{0.95}& \multicolumn{1}{c}{1.05} \\
				\hline
				\,\,$r_c^T$\,\, &&  \,\,1.47\,\, &\,\,1.63\,\, &  \,\,1.82\,\, &\,\,1.90\,\,&&\,\,1.67\,\,&\,\,1.77\,\,&\,\,1.92\,\,&\,\,2.52\,\,
				\\
				\,\,$T_+^{Max}$\,\,&\,\,&\,\,  0.0354\,\, & \,\,0.323\,\,& \,\,  0.258\,\, &\,\,  0.230\,\,&& \,\,0.0311\,\,&\,\, 0.0290\,\,&\,\,0.0270\,\,&\,\,0.256\\
				\hline
				\hline
			\end{tabular}
		\end{center}
		\caption{The maximum Hawking temperature ($T_+^{max}$) at critical radius ($r_c^{T}$) for different values of Yang-Mills charge  ($\nu$) with fixed value  ${l=20}$.}
		\label{tab:temp}
	\end{table}
\end{center}
Now, we  calculate another useful quantity associated with the black hole known as entropy $S_+$.  This black hole can be considered as a thermodynamic system only if the  quantities associated with it must obey the first-law of thermodynamics given below
\begin{eqnarray}\label{1law}
dM_+ = T_+dS_+ +\Phi\, d\nu,
\end{eqnarray}
where $\Phi$ is the potential of the black hole. It is matter of calculation only to 
derive expression of  entropy \cite{thermo,Singh:2020xju} from the expression  (\ref{1law}) at constant $\nu$ as
\begin{equation}
S_+=\pi \left(r_+^2+4\alpha\log[r_+]\right).
\end{equation}
Finally, we analyse how the background Yang-Mills field affects  the thermodynamic stability of the $4D$ EGB AdS black hole  by investigating the  heat capacity $C_+$. The stability of the black hole can be estimated from sign of the heat capacity $C_+$ as positive heat capacity reflects the stable black hole and negative reflects the unstable one. 

The heat capacity of the black hole can be estimated from the following standard relation \cite{cai}:
\be
C_+=\frac{\partial M_+}{\partial T_+}=\left(\frac{\partial M_+}{\partial r_+}\right)
\left(\frac{\partial r_+}{\partial T_+}\right).
\ee
This leads to 
\be
C_+=\frac{2\pi(r_+^2+2\alpha)^2(3r_+^4+l^2(r_+^2-\alpha-\nu^2))}{3r_+^4(r_+^2+6 \alpha)-l^2(r_+^4+2\alpha(\alpha+\nu^2)-r_+^2(5\alpha+3\nu^2))}.
\label{sh1}
\ee
From this expression, it is clearly evident that in the absence of Yang-Mills charge ($\nu$)
the heat capacity (\ref{sh1}) reduces to 
\be
C_+=\frac{2\pi(r_+^2+2\alpha)^2(3r_+^4+l^2(r_+^2-\alpha))}{3r_+^4 (r_+^2+6 \alpha)-
l^2(r_+^4+2\alpha^2-5r_+^2\alpha)},
\label{sh2}
\ee
which further  reduces to the heat capacity of the  AdS  Schwarzschild black 
hole in the limit of  $\alpha \to 0$. Moreover,  Eq. (\ref{sh1}) reduces to the  heat capacity of the 
  AdS    Schwarzschild Yang-Mills black hole when the Gauss-Bonnet coupling  switched off ($
\alpha\to 0$). 
\begin{figure*}[ht]
\begin{tabular}{c c c c}
\includegraphics[width=.5\linewidth]{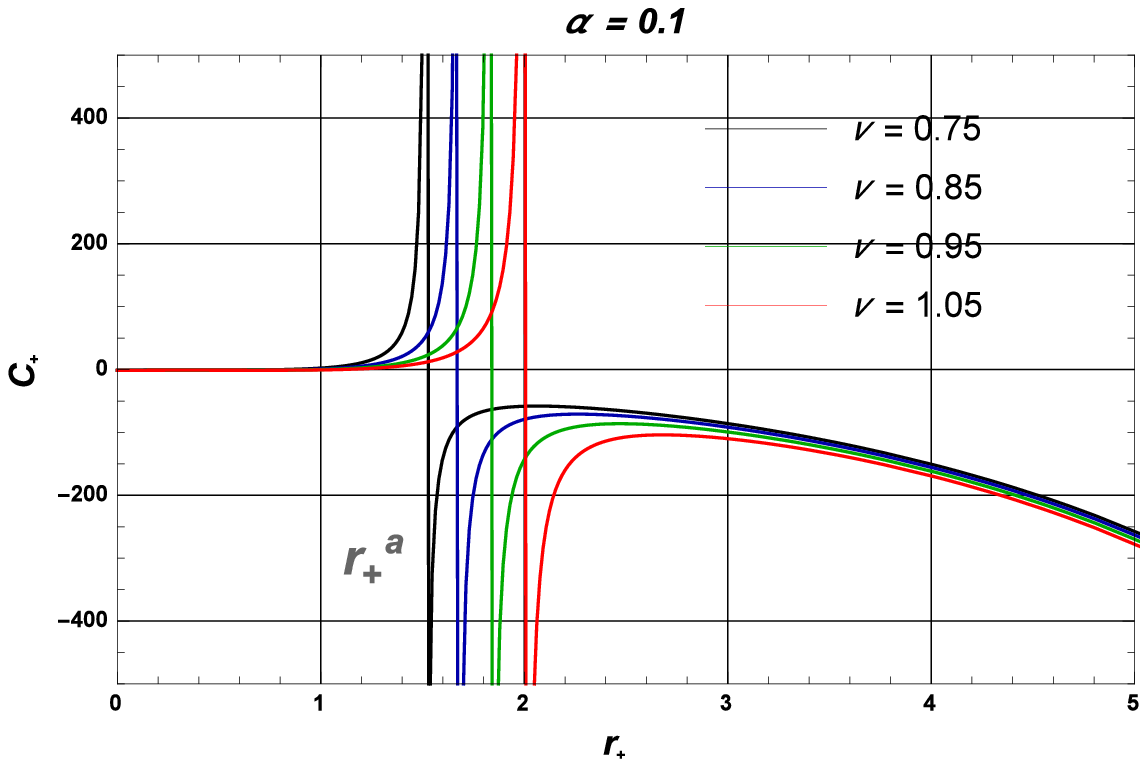}
\includegraphics[width=.5\linewidth]{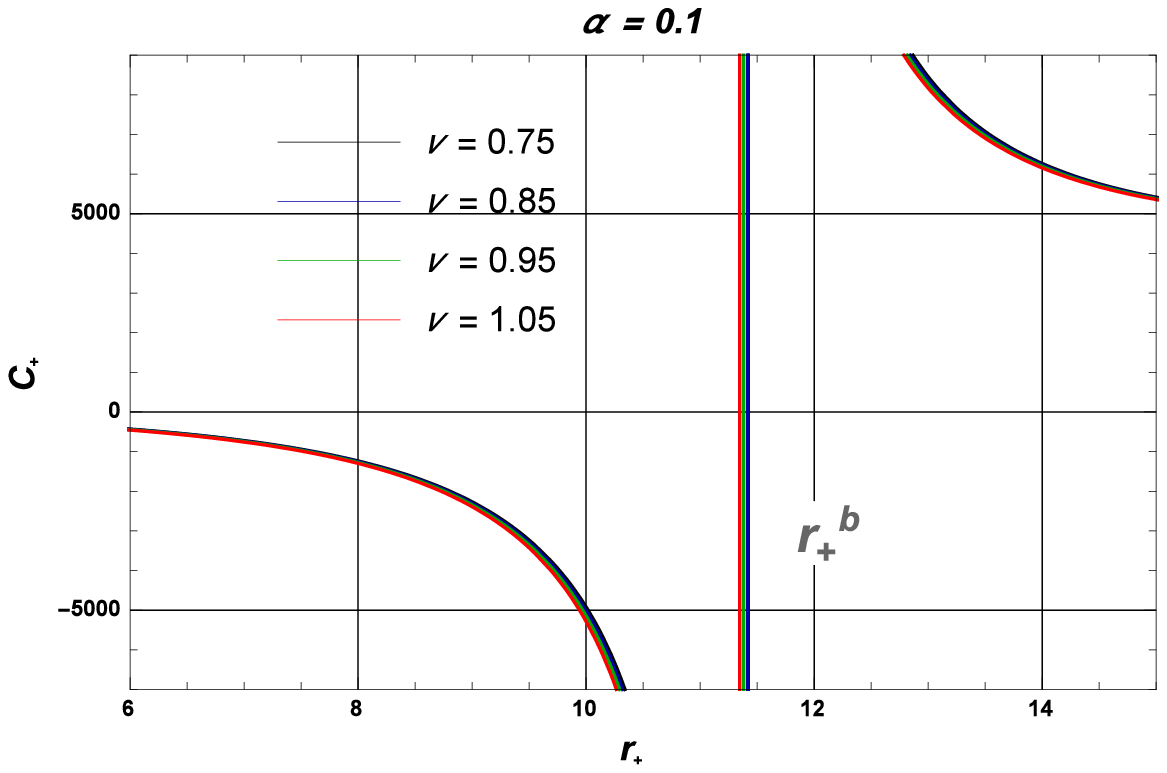}\\
\includegraphics[width=.5\linewidth]{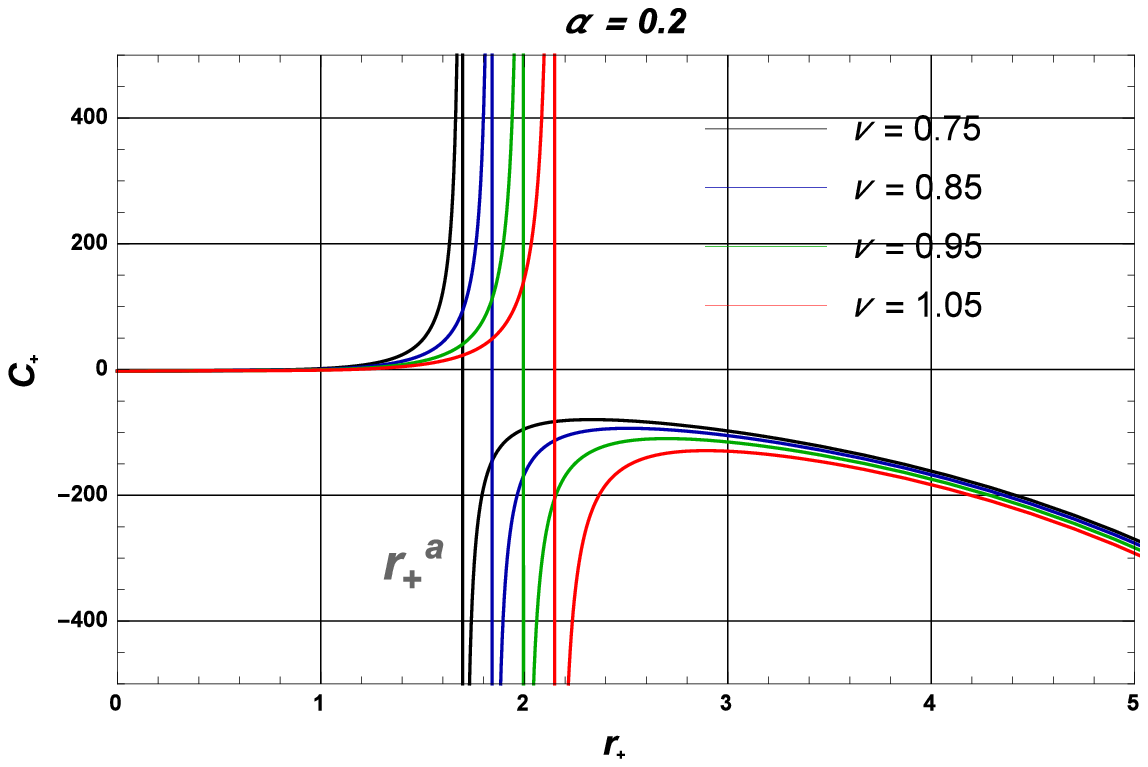}
\includegraphics[width=.5\linewidth]{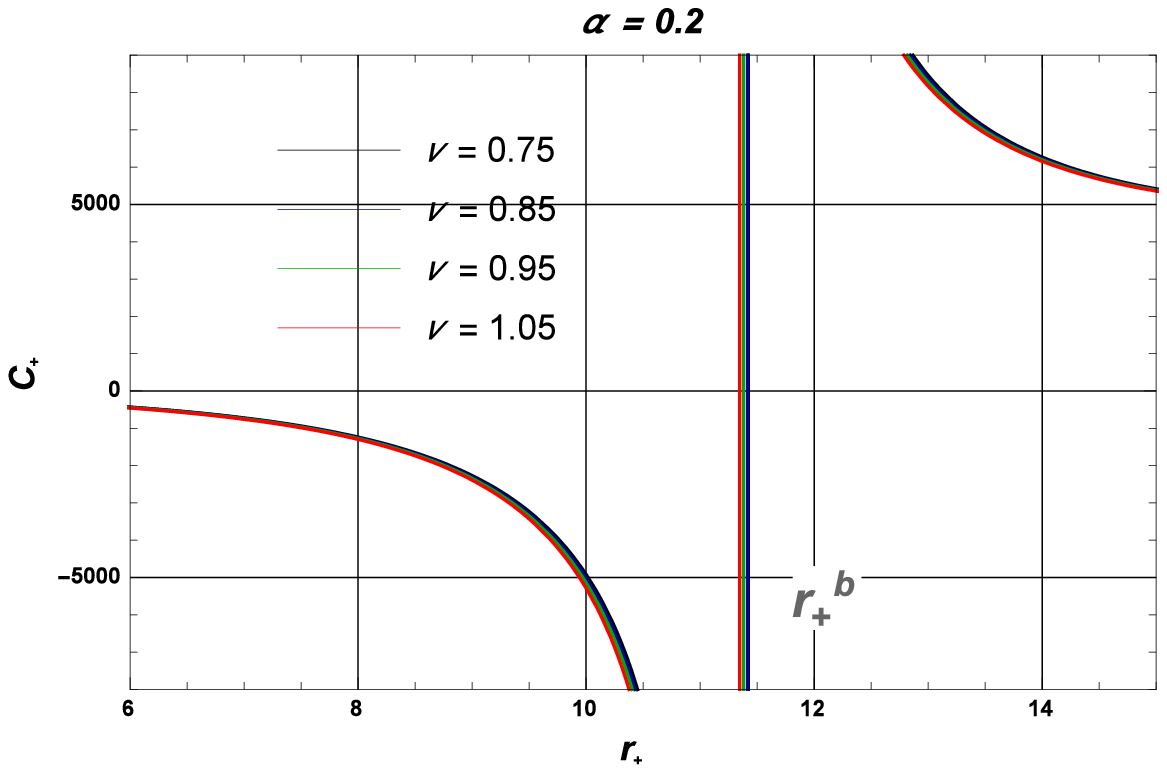}
\end{tabular}
\caption{The plot of temperature $T$ vs $r_+$ for different values of Yang-Mills charge   $\nu$  with fixed value of mass $(M=1)$ and ${l=20}$ for Gauss-Bonnet coupling parameter $\alpha=0.1$ (left) and $\alpha=0.2$ (right). }
\label{fig:6}
\end{figure*}
In order to identify the stability of the  heat capacity, we plot expression (\ref{sh1}) for different values of Yang-Mills charge $\nu$ and Gauss-Bonnet coupling $\alpha$ as shown  in Fig. \ref{fig:6}. 
From the plot, we observe that  the  heat capacity  curve  is discontinuous  where the temperature is maximum (see Fig. \ref{fig:2} and table \ref{tab:temp}).     The plot clearly shows that the heat capacity diverges at the critical points $r_+^a$ and $r_+^b$  with $r_+^a< r_+^b$  (see Fig. \ref{fig:6}).  The $4D$  AdS  EGB  Yang-Mills black hole is stable for the horizon radius $r_+<r_+^a$ and $r_+>r_+^b$ and unstable for horizon radius $r_+^a<r_+<r_+^b$. It is obvious from  this figure
that $4D$  EGB  AdS  Yang-Mills black hole undergoes phase transition twice firstly at $r_+^a$ from the stable black hole to the unstable black hole and later at $r_+^b$ from the  unstable black hole ($r_+^a<r_+<r_+^b$) to  stable black hole ($r_+>r_+^b$).  Thus, there is a phase transition occurs at $r_+=r_+^a$ and $r_+=r_+^b$  where black hole changes its phase from the stable to unstable phases and vice versa. Further, a divergence of the  heat capacity   at critical $r_+=r_c$ signals that this is  a second-order phase transition   \cite{hp,davis77}. 

The heat capacity is discontinuous at $r_+=1.47$, at which the Hawking temperature   has the maximum value $T_+=$ for $\alpha=0.1$ and $\nu=0.75$  (Fig. \ref{fig:6}). The  phase transition occurs from the higher to lower mass black holes corresponding to  positive to negative heat capacity. The critical radius $r_c$ increases along with   parameter $\alpha$ (cf. Fig. \ref{fig:6} and Table \ref{tab:temp}).

\section{Critical behaviour and phase transition  }\label{sec4}
In the extended phase space the cosmological constant $\Lambda$ is considered as the thermodynamical pressure in order to have Van der Waals like phase transition. The relation between the  thermodynamical pressure and cosmological constant is quite standard and given by
\begin{equation}
P=-\frac{\Lambda}{8\pi}=\frac{3}{8\pi l^2}.
\label{lpre}
\end{equation}
Here, units are set such that $G=\hbar= c = 1$.
The pressure  and  specific volume can be calculated as following:
\begin{equation}
P=\frac{T_+(r_+^2+2\alpha)}{2r_+^3}-\frac{r_+^2-\alpha-\nu^2}{8\pi r_+^4},\qquad v=2r_+.
\end{equation}
\begin{figure*}[ht]
\begin{tabular}{c c c c}
\includegraphics[width=.5\linewidth]{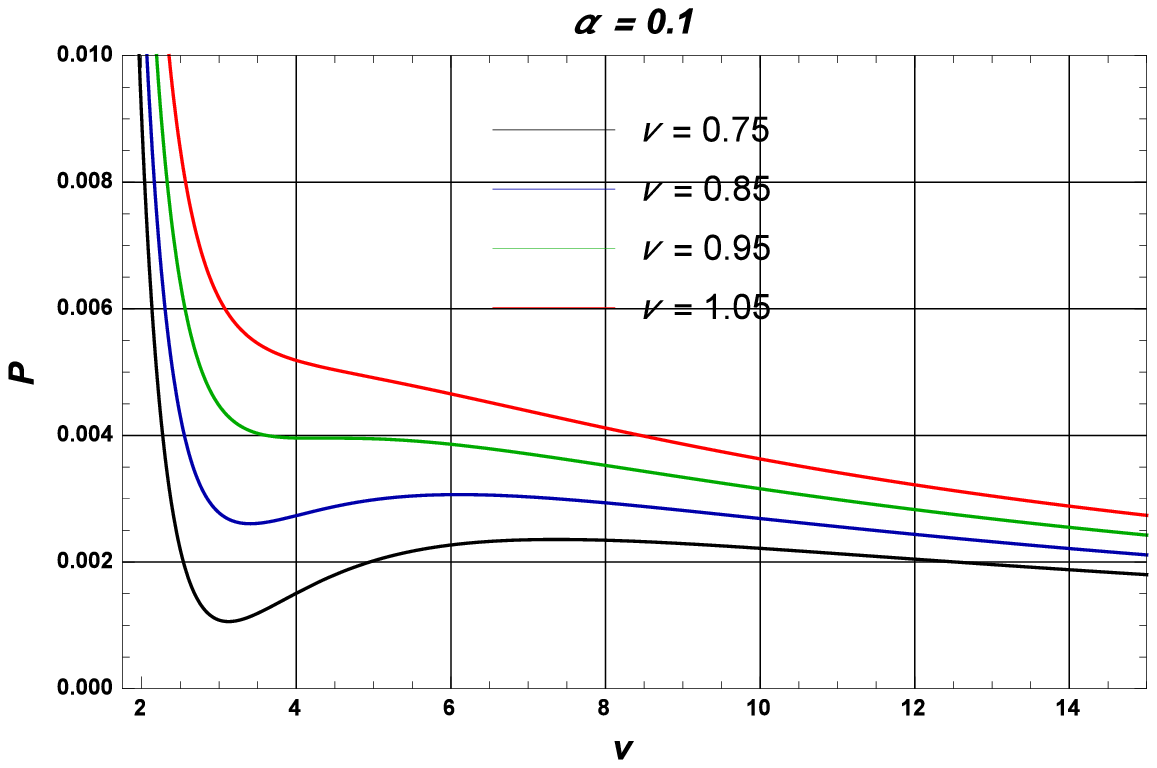}
\includegraphics[width=.5\linewidth]{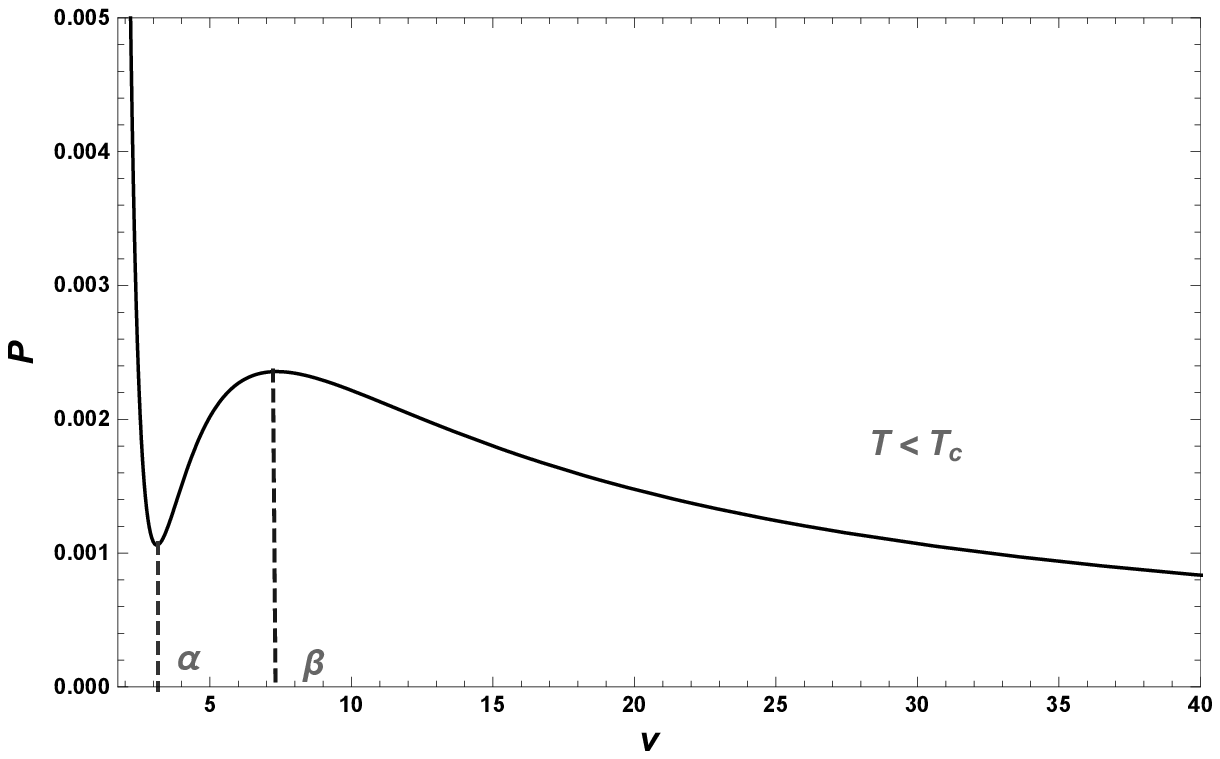}
\end{tabular}
\caption{The plot of pressure vs specific volume for $T<T_c$,  $T=T_c$ and  $T>T_c$ with $T_c=0.04038$}
\label{fig:3}
\end{figure*}
In order to study critical behaviour, we plot   a $P \--v$ diagram for the $4D$ EGB  AdS  Yang-Mills black hole as displayed in Fig. \ref{fig:3}.
Here, we  clearly see that a small-large black hole phase transition  occurs when the temperature of black hole is less than the critical temperature. 
To visualize this more clearly, a transition is directly presented on the right
panel of Fig. \ref{fig:3}.
 Notice that the $P \--v$ curve contains two stable regions and one
unstable region. These stable regions occur for $v\in [0,\alpha]$ and $v\in [\beta, \infty]$ and correspond to the small black hole region and the large black hole
region, respectively. 

The unstable region $v\in [\alpha, \beta] $ is referred as the spinodal region. In this
unstable region, the small and large black hole phase transition can coexist. Furthermore, we note that $\partial P(v,T_0)/\partial v<0$ for the stable regions, however $\partial P(v,T_0)/\partial v>0$ for the unstable region.

Mathematically, the two extreme points $\alpha$ and $\beta$ are determined by the relation
\be
\left.\frac{\partial P(v,T_0)}{\partial v}\right|_{\nu=\alpha}=\left.\frac{\partial P(v,T_0)}{\partial v}\right|_{\nu=\beta}.
\ee
Moreover, the inflection point at $\nu=\nu_0$ is determined by the relation
\be
\left.\frac{\partial^2P(v,T_0)}{\partial v^2}\right|_{T}=0.
\ee  
 The critical exponents   occur  at points of inflections in the $P-v$ diagrams
\begin{equation}
\left.\frac{\partial P_+}{\partial r_+}\right|_{T}=0,\qquad \qquad \left.\frac{\partial ^2P_+}{\partial r_+^2}\right|_{T}=0.
\label{pv5}
\end{equation}
This leads to the following values of critical radius, temperature and pressure, respectively \cite{rbm,guna}
\begin{eqnarray}
&&r_c=\sqrt{6\alpha+3\nu^2+X},\\
&&T_c=\frac{4\alpha +\nu^2+X}{2\pi (\sqrt{6\alpha+3\nu^2+X})({12\alpha+3\nu^2+X})},\\
&&P_c=\frac{52\alpha^2+\nu^2(9\nu^2+49\alpha)+X(3\nu^2+7\alpha)}{8\pi ({6\alpha+3\nu^2+X})^2({12\alpha+3\nu^2+X})},
\end{eqnarray}
where
$ X=(48\alpha^2+48\alpha\nu^2+9\nu^4)^{1/2}.$

These critical values together with relation $v_c=2r_c$  give the following universal ratio:
\be
\rho_c=\frac{P_cv_c}{T_c}=\frac{52\alpha^2+\nu^2(9\nu^2+49\alpha)+X(3\nu^2+49\alpha)}{2(6\alpha +3\nu^2+X)(4\alpha+\nu^2+X)}.
\label{uv}
\ee
In the limit of $\alpha \to 0$ the Eq. (\ref{uv}) resembles to the universal ratio $\rho_c=3/8$ \cite{rbm}. The Universal ratio of the $4D$ AdS EGB black hole $\rho_c=0.350$ is slightly smaller than the Van der Waals ratio $\rho_c=0.375$ at the small value of $\alpha$ and $\nu$  and the universal ratio increases with increasing   values of $\alpha$ and $\nu$ \cite{Hegde:2020xlv}.

\begin{figure*}[ht]
\begin{tabular}{c c c c}
\includegraphics[width=.5\linewidth]{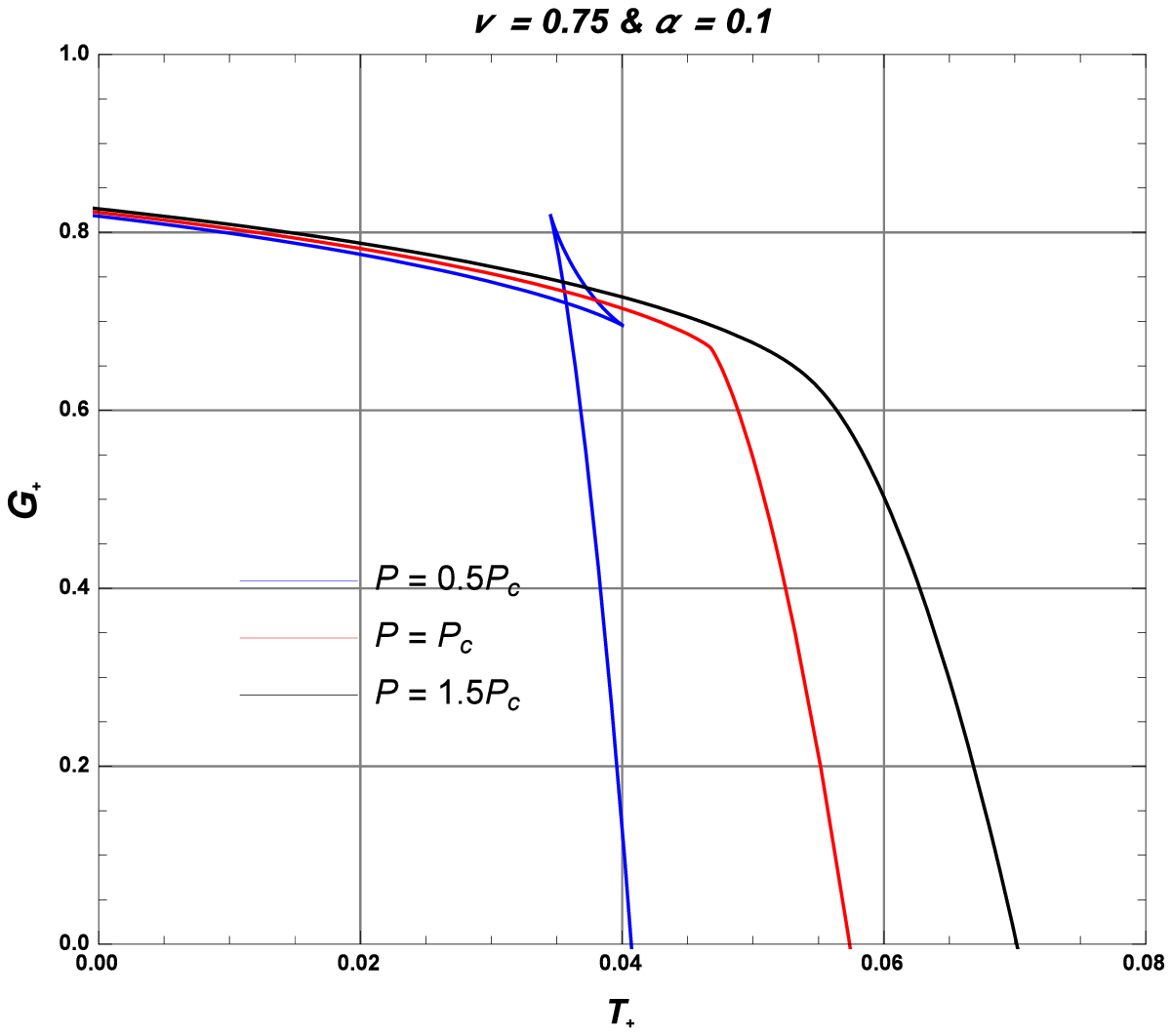}
\includegraphics[width=.5\linewidth]{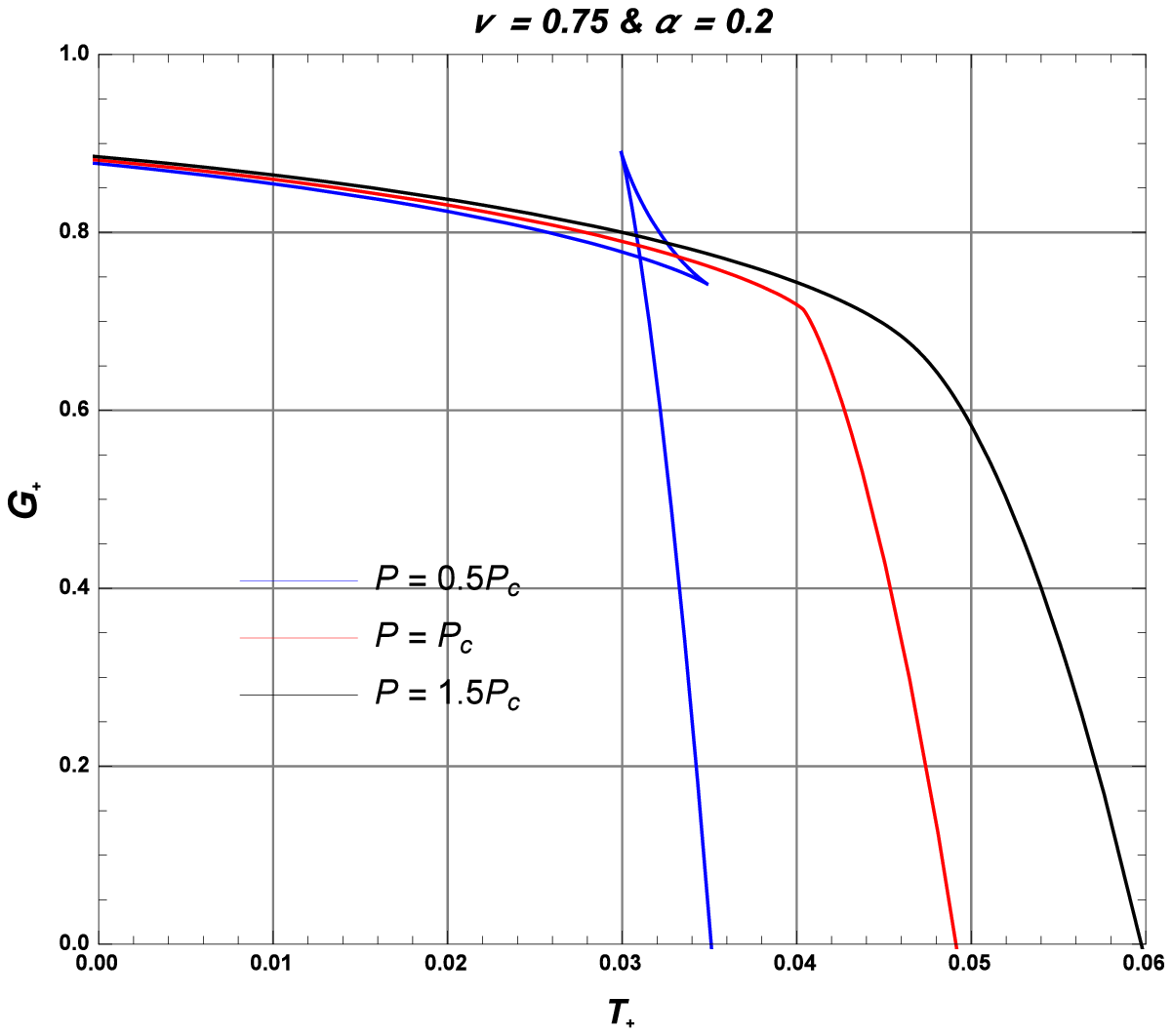}
\end{tabular}
\caption{ Gibbs free energy vs temperature of $4D$  EGB  AdS Yang-Mills  black hole for the fixed value of Yang-Mills charge $\nu$. Left: 
$\alpha =0.1$. Right: $\alpha =0.2$.}
\label{fig:4}
\end{figure*}

In order to search for the phase structure of the black hole solutions, we derive the free energy in the  canonical ensemble. 
The standard definition for Gibbs free energy  reads: $G_+=M_+-T_+S_+$  \cite{Singh:2020xju}. This  leads to 
\begin{eqnarray}
G_+&= &\frac{3r_+^2+8\pi Pr_+^4+3\alpha+3\nu^2}{6r_+} 
-  \frac{(8\pi Pr_+^4+r_+^2-\alpha-\nu^2)(r_+^2+4\alpha\log r_+)}{4r_+(r_+^2+2\alpha)}.
\end{eqnarray}
 In order to analyse the behaviour of 
Gibbs free energy, we plot the above expression in terms of temperature. 
The graph of Gibbs free energy versus temperature for $\alpha=0.1$ and  $\alpha=0.2$ with the fixed value of Yang Mills charge  $\nu$ is given plotted in Fig. \ref{fig:4}.
The Gibbs free energy  which analyses the phase transition of the black holes   analogues with the Van der Waals phase transition has the effect of Yang-Mills charge  $\nu$ and  Gauss-Bonnet coupling constant $\alpha$ on the phase structure of the system. First, we fix the Yang-Mills charge  $\nu$ and vary the Gauss-Bonnet coupling constant $\alpha$. The critical pressure and  critical temperature  increase  with critical radius (see Fig. \ref{fig:4})  with   Yang-Mills charge $\nu$.   In $G_+ - T_+$  plots the appearance of characteristic swallow tail   shows that the obtained values are critical ones where the phase transition occurs.  In  Fig. \ref{fig:4}, we can see that swallow tail shape exists when $P<P_c$   for the first order phase transition and $P=P_c$ for the second order phase transition.

\section{Critical Exponents}  \label{sec5} 

In this section, we compute the critical exponent, which is a universal property of phase transition. Usually, near the critical point, a Van der Waals like phase transition is characterized by the four critical exponents $\alpha_0$, $\beta$, $\gamma$ and $\delta$ describing the behaviour of the specific
heat,   order
parameter $\eta$,  the
isothermal compressibility $\kappa_T$ and   the
critical isotherm, respectively. These are defined as
\begin{eqnarray}
&&C_v=T\frac{\partial S}{\partial T}\propto|\tau|^{-\alpha_0},\qquad\qquad \eta=v_l-v_s \propto |\tau|^{\beta},\label{ee}\\
&&\kappa_T=-\frac{1}{v}\frac{\partial v}{\partial p}\propto|\tau|^{-\gamma},\quad\qquad |P-P_c|_{T_c}\propto |v-v_c|^{\delta},
\end{eqnarray}
where $\tau=\frac{T-T_c}{T_c}$ and $v_l$ and $v_s$ denote specific volume of large and small black holes, respectively.
 
Now, with the following definitions:
\begin{equation}
p=\frac{P}{P_c},\qquad\qquad t=\frac{T}{T_c}=\tau+1,\qquad\text{and}\qquad \lambda=\frac{v}{v_c},
\end{equation}
the equation of state translate into so-called  law of corresponding states and reads
\begin{equation}
\frac{8p}{3}=\frac{t}{\lambda}-\frac{t\alpha_0}{6\lambda^3}-\frac{}{}\left(\frac{1}{\lambda}+\frac{2\alpha_0}{\lambda}+\frac{2\nu^2}{36 \lambda^4}\right).
\label{cs}
\end{equation}
  Let us calculate the critical exponents predicted by the corresponding states (\ref{cs}). We define
\begin{equation}
\qquad\qquad \omega=\frac{v-v_c}{v_c}=\lambda-1.
\label{es1}
\end{equation}
To find the critical exponents we expend equation of state near the critical points as following:
\begin{equation}
p=1+a_{10}\tau-a_{11}\tau\omega-a_{30}\omega^3+{\cal O}(\tau\omega^2,\omega^4),
\label{series}
\end{equation}
In the absence of Gauss-Bonnet coefficient,  the above coefficients resemble  to the Reissner-Nordstr\"om black hole. Differentiating the Eq. (\ref{series}) with respect to $\omega$ for fixed value of $t$, we get
\begin{equation}
dP=P_c\left(a_{11}\tau+a_{30}\omega^2\right)d\omega.
\label{series1}
\end{equation}
If the black hole undergoes the phase transition from small to large one keeping temperature and pressure
constant while changing the thermodynamic volume from $\omega_s$ to $\omega_l$, the equation of state always holds,
i.e.,
\begin{eqnarray}
 p=1+a_{10}\tau-a_{11}\tau\omega_l-a_{30}\omega_l^3=1+a_{10}\tau-a_{11}\tau\omega_s-a_{30}\omega_s^3.
\end{eqnarray} 
Moreover, during the phase transition, the Maxwell's area law also holds, i.e.,
\begin{eqnarray}
 \int_{\omega_l}^{\omega_s}\omega dP=\omega\left(a_{11}\tau+ a_{30}\omega^2\right)dP=0.
\end{eqnarray} 
Solving for the above equations, we obtain a non-trivial solution
\begin{equation}
\omega_l=-\omega_s=\sqrt{-\frac{a_{11}\tau}{a_{30}}}=2\sqrt{-\tau}.
\end{equation}
Hence we have the order parameter $\eta$ as
\begin{equation}\label{e}
\eta=v_c(\omega_l-\omega_s)=\sqrt{-\frac{a_{11}\tau}{a_{30}}}\propto\sqrt{-t}=4v_c\sqrt{-\tau}.
\end{equation}
The relations (\ref{ee}) and (\ref{e}) confirm that 
$\beta =1/2$.

The isothermal compressibility $\gamma$ is calculated by
\begin{equation}
\kappa_T=-\left.\frac{1}{P_c(1+\omega)}\frac{\partial \omega}{\partial P}\right|_T\propto\frac{1}{6P_c}\frac{1}{\tau}. 
\end{equation}
This implies that $\gamma=1$.

The critical isotherm at $\tau=0$ is estimated   from Eq. (\ref{series}) as
$\delta =3$ as $p-1=-\frac{4}{81}(1-\alpha_0)\omega^3$.
Consequently, the four critical exponents are identified as
\begin{equation}
\alpha_0=0,\qquad\beta=\frac{1}{2},\qquad \gamma=1\qquad\text{and}\qquad\delta=3.
\end{equation}
Obviously the entropy is not dependent of temperature and hence the critical exponent $\alpha_0=0$. These critical exponents resembles to the Van der Walls fluid  and these critical exponents satisfy the following thermodynamics scaling laws
\begin{eqnarray}
&& \alpha_0+2\beta+\gamma=2,\qquad \alpha_0+\beta(\delta+1)=2,\qquad(1-\alpha_0)\delta=1+(2-\alpha_0),\\
&&\gamma(\delta+1)=(2-\alpha_0)(\delta-1)\qquad \text{and} \qquad \gamma=\beta(\delta-1).
\end{eqnarray}
We conclude that the thermodynamics exponents associated with the  AdS EGB black hole with Yang-Mills field  in the extended phase  agree with the  mean field theory prediction \cite{rbm,guna,lala}.

Charged (Abelian) black hole solution with some thermal properties have been studied so far for the $4D$ AdS  EGB model  \cite{fran, Singh:2020xju,Hegde:2020xlv}. Here, we have generalized the black hole solution for  $4D$ AdS  EGB gravity model in presence of Yang-Mills (non-Abelian) charge. In fact, we explore the  critical behaviour, phase transition and the critical exponents  also  along with thermodynamics of the model.   
 With an appropriate limiting case, one can recover the solutions of   aforementioned references.

\section{Discussions and Conclusions}\label{sec7}
We have found  an  exact solution for $4D$ EGB black hole in the presence of Yang-Mills field which is a generalization of the solution obtained by Glavan and Lin  \cite{gla}. Also, one recovers the Schwarzschild  AdS solution   for $\alpha\rightarrow 0$, $\nu\rightarrow   0$ and $l \rightarrow \infty$.   
 Furthermore, we have  investigated the thermodynamics behaviour of the solution. Here, we have observed that the thermodynamics quantities get modification due to the presence of Yang-Mills charge $\nu$. However,  the entropy of the black hole remains unaffected by the Yang-Mills charge and  turns out to be the Bekenstein-Hawking area-law for the  novel $4D$ EGB black hole. In order to discuss the stability of the black hole we derive the heat capacity. The heat capacity $C_+$ diverges at a critical radius $r=r_c$, where the temperature has a maximum and $C_+ > 0$ for $r_+ < r_c$ allowing the black hole to become thermodynamically stable.

We also examined the $P\--v$ criticality and phase structure of the $4D$ AdS  EGB black hole solution with Yang-Mills charge. We have found  that a small-large black hole phase transition  occurs for the case where temperature is less than the critical temperature, i.e., $T< T_c$. The critical exponents are also calculated   which   increase   with the Yang-Mills charge   $\nu$ and decrease with the GB coupling constant $\alpha$. It will be interesting to study the effect of  thermal and geometric fluctuation 
on the thermodynamics of this black hole solution. This is the subject of future investigation.


\appendix
\section{The equations of motion for EGB gravity with Yang-Mills field}\label{ap}
The action of the $D$-dimensional  EGB gravity in the presence of  Yang-Mills field is given by
\begin{eqnarray}
S&=&\frac{1}{2}\int d^{D}x\sqrt{-g}\left[ {\cal R}-2\Lambda +\frac{\alpha}{D-4} {\cal L_{GB}}-{\cal F}_{YM} \right],
\label{action}
\end{eqnarray}
where $ \mathcal{L_{GB}}$  is Lagrangian density for EGB gravity and  has following expression: $$ \mathcal{L_{GB}}={\cal R}_{\mu \nu \gamma \delta }{\cal R}^{\mu \nu \gamma \delta}-4{\cal R}_{\mu \nu }{\cal R}^{\mu \nu }+{\cal R}^{2}.$$  The equation of motion  for metric tensor $g_{\mu\nu}$ and the electromagnetic potential ${\cal A}_{\mu}^{(a)}$ are given, respectively, by
\begin{eqnarray}
&&G_{\mu \nu }+\Lambda g_{\mu \nu }+H_{\mu \nu }=T_{\mu\nu}^{YM},\\
&&D_{\mu}F^{(a)\mu\nu}=0,\label{3}
\label{Field equation}
\end{eqnarray}
 where  
\begin{eqnarray}
G_{\mu\nu}&=&{\cal R}_{\mu\nu}-\frac{1}{2}g_{\mu\nu}{\cal R},\\
H_{\mu \nu }& =&-\frac{\alpha }{2}\left[ 8{\cal R}^{\rho \sigma }{\cal R}_{\mu \rho \nu
\sigma }-4{\cal R}_{\mu }^{\rho \sigma \lambda }{\cal R}_{\nu \rho \sigma \lambda
}-4{\cal R}{\cal R}_{\mu \nu }+8{\cal R}_{\mu \lambda }{\cal R}_{\nu }^{\lambda }\right.  \nn  \\
&+&\,\,\left. g_{\mu \nu }\left( {\cal R}_{\rho\sigma \gamma \delta }{\cal R}^{\rho\sigma \gamma
\delta }-4{\cal R}_{\rho\sigma }{\cal R}^{\rho\sigma }+{\cal R}^{2}\right) \right],\nonumber\\
 T_{\mu\nu}^{YM}&=&-\frac{1}{2}g_{\mu\nu}F_{\rho\sigma}^{(a)}F^{(a)\rho\sigma}+2F_{\mu\sigma}^{(a)}F^{(a)\sigma}_{\ \nu}.
\end{eqnarray}
Here ${\cal R}$, ${\cal R}_{\mu\nu}$ and ${\cal R}_{\mu\nu\gamma\delta}$  the $D = d + 1$ dimensional Ricci scalar, Ricci tensor and Riemann tensor, respectively.

\end{document}